\documentclass[aps,preprint,showpacs]{revtex4}
\usepackage{amsmath}
\newcommand{\be}{\begin{equation}}
\newcommand{\ee}{\end{equation}}
\newcommand{\bea}{\begin{eqnarray}}
\newcommand{\eea}{\end{eqnarray}} 
\newcommand{\ba}{\begin{array}}
\newcommand{\ea}{\end{array}}

\newcommand{\bb}{\bibitem}

\begin{document}

\title{\bf Callan-Symanzik method for $m$-axial Lifshitz points}
\author{Paulo R. S. Carvalho\footnote{e-mail:renato@df.ufpe.br} and 
Marcelo M. Leite\footnote{e-mail:mleite@df.ufpe.br}}
\affiliation{{\it Laborat\'orio de F\'\i sica Te\'orica e Computacional, Departamento de F\'\i sica,\\ Universidade Federal de Pernambuco,\\
50670-901, Recife, PE, Brazil}}

\vspace{0.2cm}
\begin{abstract}
{\it We introduce the Callan-Symanzik method in the description of anisotropic 
as well as isotropic Lifshitz critical behaviors. Renormalized perturbation 
theories are defined by normalization conditions with nonvanishing masses and 
at zero external momenta. The orthogonal approximation is employed to obtain 
the critical indices $\eta_{L2}$, $\nu_{L2}$, $\eta_{L4}$ and $\nu_{L4}$ 
diagramatically at least up to two-loop order in the anisotropic 
criticalities. This approximation is also utilized to compute the exponents 
$\eta_{L4}$ and $\nu_{L4}$ in the isotropic case. Furthermore, we compute 
those exponents exactly for the isotropic behaviors at the same loop order. 
The results obtained for all exponents are in perfect agreement with those 
previously derived in the massless theories renormalized at nonzero external 
momenta.}
\end{abstract}

\vspace{1cm}

\maketitle

\newpage
\section{Introduction}

Renormalized perturbative expansion of a $\lambda\phi^{4}$ field theory is the 
natural mathematical setting in statistical mechanics to determine the 
critical properties taking place in second order phase transitions. For 
ordinary critical systems, the theory is defined on a $d$-dimensional 
Euclidean space and has quadratic kinetic terms in momenta space, as usual 
in quantum field theory. These criticalities can be generalized by adding 
certain combinations of $m$ quartic derivatives of the field to the kinetic 
term of the bare Lagrangian density. In fact, when the $m$ coefficients of 
the usual quadratic derivatives of the field vanish, the Lagrangian density 
describes the so-called $m$-axial Lifshitz critical 
behaviors\cite{Ho-Lu-Sh,Ho,Se1}. Physically, the bare Lagrangian describes an 
Ising model with ferromagnetic exchange forces between nearest neighbor spins 
as well as antiferromagnetic interactions between second neighbors along $m$ 
space directions. The competition among ferro- and antiferromagnetic couplings 
of the spins in the model induces the vanishing of the $m$ terms in the 
quadratic derivatives of the field keeping, however, the quartic derivatives 
along the competing directions. The $m$ higher derivative terms will generate 
space anisotropy whenever $m < d$, with two inequivalent subspaces, whereas 
if $m=d$ there is only one isotropic space. 

From the quantum field-theoretic side, interest in the model stems from the 
fact that it has a natural connection with Lorentz symmetry breaking for 
fields propagating in specific backgrounds in the long-distance limit. Indeed, 
a recent proposal postulates the existence of a physical fluid originating 
from a scalar field whose vacuum expectation value changes with a constant 
velocity \cite{Arka1}. This background breaks the time diffeomorphism symmetry 
generating a type of Goldstone boson for this breaking of Lorentz invariance, 
the ghost condensate (analogous to the Goldstone bosons which arise from 
spontaneous breaking of internal/gauge symmetries). It mimics a type 
of dark energy, whose low-energy effective Lagrangian in flat spacetime has 
kinetic terms with quadratic time derivatives and quartic space derivatives. 
The ghost condensate mixes with gravity in a kind of Higgs mechanism giving 
rise to a nontrivial modification of gravity in the infrared \cite{Arka2}. 
The competing interactions between attractive and repulsive components of 
gravity at large scales manifest themselves in the absence of quadratic 
space derivatives of the ghost condensate in the kinetic Lagrangian implying 
that this flat spacetime is not the conventional Minkowski space. After a Wick 
rotation in time for this quantum field theory, the kinetic Lagrangian for the 
ghost condensate ($\pi$) can be identified with that for the order parameter 
field (magnetization $\phi$) in the $m$-axial Lifshitz critical behavior for 
$m=3$. Therefore, figuring out the perturbative structure 
of Lifshitz points is worthwhile in order to gain a better comprehension 
of this kind of Lorentz breaking quantum field theory as well 
as the large distance effects on the gravity sector in this model.
   
Renormalization group and $\epsilon$-expansion ideas 
($\epsilon=4-d)$ were originally introduced to provide the perturbative 
determination of critical exponents using diagramatic methods in momenta 
space \cite{Wilson}. A reformulation of this method was introduced 
to compute the critical exponents in a massless theory renormalized at 
nonzero external momenta\cite{BLZ1,amit}. The associated renormalization 
group equation along with further developments in evaluating Feynman 
diagrams have been adapted to study critical exponents of the Lifshitz type 
when more than one characteristic length scale are present \cite{comment1}. 
Critical exponents of Lifshitz points have previously been 
obtained applying these ideas in massless theories renormalized at nonzero 
external momenta\cite{Lei1,DS}.   

An alternative to the above methods is to formulate the problem in a massive 
theory. This permits to comprehend under what conditions the scale invariance 
of the solution to the renormalization group equation for the renormalized 
1PI vertex parts is guaranteed. In addition, this formulation is convenient 
to treat renormalization issues of arbitrary quantum field theories and it 
is a healthy test for the universality hypothesis, which guarantees that 
universal quantities (e.g., the critical exponents) do not depend on the 
scheme of renormalization. For ordinary critical systems, the asymptotic 
scale invariance of the theory in $d=4-\epsilon$ was established more 
rigorously after the first developments Ref.\cite{Wilson} through the use 
of Callan-Symanzik equations \cite{C,Z} in the explicit computation of the 
critical exponents for a massive theory\cite{BLZ2}. However, the same approach 
has not been applied for investigating more general critical systems so far. 
In fact we would like to know if such a method could be adapted to critical 
systems exhibiting more than one characteristic length scale, such as the 
$m$-fold Lifshitz critical behavior. It is the purpose of the present work 
to show that such a mission can be accomplished, following closely the 
arguments 
put forth in \cite{BLZ2}. Here, we derive the critical exponents $\eta$ and 
$\nu$ in the massive case by introducing an appropriate Callan-Symanzik 
formalism with two independent mass scales in the bare Lagrangian 
corresponding to the correlation lengths  $\xi_{L2}$ and $\xi_{L4}$ associated 
to the anisotropic critical behaviors. The massive theories are renormalized at zero external momenta. 
A similar treatment will be shown to be valid for the isotropic behaviors 
with only one mass scale. Consequently, we derive the critical exponents 
characterizing the anomalous dimensions of the field $\phi$ and composite 
operator $\phi^{2}$, respectively, within an $\epsilon_{L}$-expansion at 
least at two-loop order using Feynman's path integral method in momentum 
space.

The anisotropic diagrams are calculated using the orthogonal approximation 
introduced in \cite{Lei1}. The isotropic diagrams are more flexible and can 
be calculated simultaneously using the orthogonal approximation as well as 
exactly. The exact computation was recently demonstrated in the massless 
theory \cite{Lei2} and shall be calculated here in the massive theory. The 
exponents computed diagrammatically in the present work agree with those 
obtained in the massless theory renormalized at nonvanishing external momenta 
as expected.   

The organization of the paper is as follows. In Sec. II we present the 
normalization conditions for the anisotropic and isotropic Lifshitz critical 
behaviors in the massive case. The Callan-Symanzik equations and Wilson 
functions are defined in Sec. III. The critical exponents for the anisotropic 
cases are presented in Sec. IV. The computation of critical indices for the 
isotropic cases is the subject of Sec. V. The conclusions and perspectives 
of the present work are discussed in Sec VI. The relevant Feynman integrals 
are calculated in the appendixes. Appendix A contains the information 
concerning the computations involved in the anisotropic behavior using an 
approximation to resolve the Feynman integrals. The isotropic integrals are 
computed using the same approximation in Appendix B. Finally, the exact 
isotropic integrals computed exactly are summarized in Appendix C.   

\section{Normalization conditions for the massive theories}

Before setting the normalization conditions let us recall some basic facts 
about the Lifshitz critical behaviors. The  bare Lagrangian density is that 
of a self-interacting scalar field given by\cite{Lei1} :
\begin{equation}\label{1}
L = \frac{1}{2}|\bigtriangledown_{m}^{2} \phi_0\,|^{2} +
\frac{1}{2}|\bigtriangledown_{(d-m)} \phi_0\,|^{2} +
\delta_0  \frac{1}{2}|\bigtriangledown_{m} \phi_0\,|^{2}
+ \frac{1}{2} \mu_{\tau}^{2\tau}\phi_0^{2} + \frac{1}{4!}\lambda_{0\tau}\phi_0^{4} .
\end{equation}

The Lifshitz point, where $\mu_{\tau}=0$ (for the temperature fixed at the 
Lifshitz value $T=T_{L}$) and $\delta_{0}=0$ is the multicritical 
point of interest in this framework. Consider the anisotropic criticalities. 
The first term in the above expression should be multiplied by a constant in 
order to make sense on dimensional grounds. Fortunately, the condition 
$\delta_{0}=0$, which will be implemented in all the subsequent discussion, 
permits the dimensional redefinition of momentum scales along 
the competing $m$-dimensional axes with quartic momentum dependence. They have 
half the value of a genuine momentum scale and we can get rid of the constant 
in front of this unusual kinetic term. In the above Lagrangian we introduce 
two independent bare masses in order to perform the appropriate scale 
transformations consistent with the two independent correlation lengths 
characterizing these situations. They have different powers in order to 
emphasize that the momenta subspaces are inequivalent. Thus, the bare mass 
$\mu_{2}$, as well as its renormalized counterpart $m_{2}$ have half the 
value of a genuine mass scale. These distinctions do bring new insights for 
it is convenient that mass and momentum have the same canonical dimension in 
each subspace separately. In addition, two independent cutoffs are 
needed to provide independent flows in the parameter space as far as 
the masses are concerned. Together with two coupling constants chacterizing 
each subspace, we have an apparent overcounting which allows the 
renormalization group transformations to be treated independently in the 
inequivalent subspaces. In the isotropic critical behavior there are some 
differences: the second term in the Lagrangian is absent and there is only one 
type of bare (and renormalized) mass(es).  

It is interesting to consider the renormalized theories that can be 
constructed out of the Lagrangian (1). To do so, we shall follow the 
conventions addopted in \cite{Lei1}. The reader should also consult 
Ref.\cite{amit} for further notation details.

Let us turn our attention to the anisotropic cases. The two subspaces, 
namely one with a quadratic dependence on the momenta and the 
other with a quartic dependence on the momenta are orthogonal to each other 
and can be treated independently. In fact, we can 
define two sets of normalization conditions associated to either subspace as 
follows. The quadratic subspace is defined by $\mu_{2}, \lambda_{2}=0$, 
$\mu_{1}, \lambda_{1} \neq 0$. The latter induce renormalized quantities 
$m_{1}$ and $g_{1}$. Since no infrared divergence takes place in massive 
theories, we can renormalize the theory at zero external momenta. The 
one-particle irreducible(1PI) renormalized vertex parts are defined 
through the following normalization conditions:
\begin{subequations}
\begin{eqnarray}
&& \Gamma_{R(1)}^{(2)}(0,m_{1}, g_{1}) = m_{1}^{2}, \\
&& \frac{\partial\Gamma_{R(1)}^{(2)}(p, m_{1}, g_{1})}{\partial p^{2}}|_{p^{2}=0} = 1, \\
&& \Gamma_{R(1)}^{(4)}(0, m_{1}, g_{1}) = g_{1}  , \\
&& \Gamma_{R(1)}^{(2,1)}(0, 0, m_{1}, g_{1}) = 1 , \\
&& \Gamma_{R(1)}^{(0,2)}(0, m_{1}, g_{1}) = 0 .
\end{eqnarray}
\end{subequations} 
 The subscript 1 makes explicit reference to the quadratic or noncompeting 
subspace. The most general vertex parts at this subspace have nonvanishing 
external momenta along the quadratic directions as well as zero external 
momenta components along the quartic directions. 

Similarly, the definition of renormalized theories along the quartic or 
competing directions can be performed by setting first all the external 
momenta along the noncompeting directions equal to zero and taking 
$\mu_{1}, \lambda_{1}=0$ with $\mu_{2}, \lambda_{2} \neq 0$. The resulting 
1PI renormalized parts in the competing subspace can be rendered finite by 
using the following set of normalization conditions:
\begin{subequations}
\begin{eqnarray}
&& \Gamma_{R(2)}^{(2)}(0, m_{2}, g_{2}) = m_{2}^{4}, \\
&& \frac{\partial\Gamma_{R(2)}^{(2)}(p, m_{2}, g_{2})}{\partial p^{4}}|_{p^{4}=0} \
= 1, \\
&& \Gamma_{R(2)}^{(4)}(0, m_{2}, g_{2}) = g_{2}  , \\
&& \Gamma_{R(2)}^{(2,1)}(0, 0, m_{2}, g_{2}) = 1 , \\
&& \Gamma_{R(2)}^{(0,2)}(0, m_{2}, g_{2}) = 0 .
\end{eqnarray}
\end{subequations} 
The subscript 2 refers to the competing subspace. Therefore, the competing 
and noncompeting subspaces are effectively separated out. In order to be 
completely similar to the massless case normalization conditions devised in 
Ref.\cite{Lei1}, we can fix the mass scale of the two-point functions in 
either subspace by the choices $m_{\tau}^{2\tau}=1$. Furthermore, the 
isotropic critical behavior can be defined using only the second set of 
normalization conditions from the anisotropic behaviors, with different 
values of renormalized mass and coupling constants, say, $m_{3}, g_{3}$. 
Next, we apply these conditions to the discussion of the Callan-Symanzik 
equations for the anisotropic and isotropic $m$-axial critical behaviors.   
  
\section{The Callan-Symanzik equations}

Let the label  $\tau = 1,2$ refer to the different external momenta scales
involved in the general Lifshitz critical behavior, as discussed above for
different normalization conditions in the anisotropic and isotropic cases. 
In terms of the bare quantities, the renormalized 1PI vertex parts are 
defined by  
\begin{eqnarray}
\Gamma_{R(\tau)}^{(N,L)} (p_{i (\tau)}, Q_{i(\tau)}, g_{\tau}, m_{\tau})
&=& Z_{\phi (\tau)}^{\frac{N}{2}} Z_{\phi^{2} (\tau)}^{L}
(\Gamma^{(N,L)} (p_{i (\tau)}, Q_{i (\tau)}, \lambda_{\tau}, \mu_{\tau}, \Lambda_{\tau})\\ \nonumber
&& - \delta_{N,0} \delta_{L,2}
\Gamma^{(0,2)}_{(\tau)} (Q_{(\tau)}, Q_{(\tau)}, \lambda_{\tau}, \mu_{\tau}, \Lambda_{\tau})|_{Q^{2}_{(\tau)} = 0})
\end{eqnarray}
where $p_{i (\tau)}$ ($i=1,...,N$) are the external momenta associated to
the vertex functions $\Gamma_{R(\tau)}^{(N,L)}$ with $N$ external legs, 
$Q_{i (\tau)}$ ($i=1,...,L$) are the external momenta associated to the
$L$ insertions of $\phi^{2}$ operators and $\Lambda_{\tau}$ are cutoffs 
characterizing each inequivalent subspace. In terms of the dimensionless 
couplings $u_\tau$, the renormalized and bare coupling constants are given, 
respectively, by 
$g_{\tau} = u_{\tau} (m_{\tau}^{2 \tau})^{\frac{\epsilon_{L}}{2}}$,
and $ \lambda_{\tau} =  u_{0 \tau} 
(m_{\tau}^{2 \tau})^{\frac{\epsilon_{L}}{2}}$,
where $\epsilon_{L}= 4 + \frac{m}{2} - d$. In addition, we can write all the 
renormalization functions in terms of $u_\tau$.

By expanding the dimensionless bare 
coupling constants $u_{o \tau}$ and the renormalization functions 
$Z_{\phi (\tau)}$, $\bar{Z}_{\phi^{2} (\tau)} = Z_{\phi (\tau)} Z_{\phi^{2} (\tau)}$ in terms of the dimensionless renormalized couplings $u_{\tau}$ up 
to two-loop order as
\begin{subequations}
\begin{eqnarray}
&& u_{o \tau} = u_{\tau} (1 + a_{1 \tau} u_{\tau} + a_{2 \tau} u_{\tau}^{2}) ,\\
&& Z_{\phi (\tau)} = 1 + b_{2 \tau} u_{\tau}^{2} + b_{3 \tau} u_{\tau}^{3} ,\\
&& \bar{Z}_{\phi^{2} (\tau)} = 1 + c_{1 \tau} u_{\tau} + c_{2 \tau} u_{\tau}^{2} ,
\end{eqnarray}
\end{subequations}
we can unravel the renormalization structure of the Callan-Symanzik (CS) 
equations as they arise in the two situations. Let us now make a distinction 
in the two cases by considering separately the CS equations for anisotropic 
and isotropic spaces.

\subsection {Anisotropic}

In spite of having two independent mass scales, the asymptotic 
behaviors of the anisotropic cases at the critical dimension 
$d_{c}= 4 + \frac{m}{2}$ proceed along the same lines as the 
ordinary critical behavior at 4 dimensions described by a $\lambda\phi^{4}$ 
field theory. Therefore, we shall limit ourselves to the situation where 
$d = 4 + \frac{m}{2} - \epsilon_{L}$.

Before deriving explicitly the equations which relate theories renormalized 
at different mass scales, let us recall some useful definitions for certain quantities in terms of dimensionless coupling constants. The $beta$-functions 
and Wilson functions for the anisotropic behaviors are defined by:
\begin{subequations}
\begin{eqnarray}
&& \beta_{\tau} = (m_{\tau}\frac{\partial u_{\tau}}{\partial m_{\tau}}), \\
&& \gamma_{\phi (\tau)}(u_{\tau})  = \beta_{\tau}
\frac{\partial ln Z_{\phi (\tau)}}{\partial u_{\tau}}\\
&& \bar{\gamma}_{\phi^{2} (\tau)}(u_{\tau}) = - \beta_{\tau}
\frac{\partial ln {\bar{Z}_{\phi^{2} (\tau)}}}{\partial u_{\tau}}
\end{eqnarray}
\end{subequations}
are calculated at fixed bare coupling $\lambda_{\tau}$. Since the theory is 
massive, scale invariance will be achieved in the limit where the cutoffs 
$\Lambda_{\tau}$ go to infinity. The $\beta_{\tau}$-functions can be cast in 
a more useful form in terms of dimensionless quantities, namely,

\begin{equation}
\beta_{\tau} = - \tau \epsilon_{L}(\frac{\partial ln u_{0 \tau}}{\partial u_{\tau}})^{-1}.
\end{equation}
By differentiating Eq.(4) with respect to $lnm_{\tau}$ we find 
\begin{eqnarray}
&&(m_{\tau} \frac{\partial}{\partial m_{\tau}} +
\beta_{\tau}\frac{\partial}{\partial g_{\tau}}
- \frac{1}{2} N \gamma_{\phi (\tau)}(u_{\tau}) + L \gamma_{\phi^{2} (\tau)}(u_{\tau}))
\Gamma_{R(\tau)}^{(N,L)} (p_{i (\tau)}, Q_{i (\tau)}, g_{\tau},
m_{\tau})  \nonumber \\ 
&& - \delta_{N,0} \delta_{L,2} (\kappa_{\tau}^{-2 \tau})^{\frac{\epsilon_{L}}{2}} B_{\tau}= m_{\tau}^{2 \tau}(2 \tau - \gamma_{\phi (\tau)})\Gamma_{R(\tau)}^{(N,L+1)} (p_{i (\tau)}, Q_{i (\tau)},0, g_{\tau},m_{\tau})
\end{eqnarray}
where $B_{\tau}$ is a constant used to
renormalize $\Gamma_{R (\tau)}^{(0,2)}$ and 
$\gamma_{\phi^{2} (\tau)}(u_{\tau})= - \beta_{\tau}
\frac{\partial ln {Z_{\phi^{2} (\tau)}}}{\partial u_{\tau}}$ . The last equations are the 
Callan-Symanzik equations for anisotropic vertex parts with arbitrary 
composite operators. The right hand side can be asymptotically neglected 
order by order in perturbation theory. Consequently, the 
Callan-Symanzik (CS) equations have asymptotically the same 
form as the renormalization group (RG) equations previously derived 
in Ref.\cite{Lei1}. 

For the sake of simplicity, consider the vertex parts without composite 
operators with $L=0$. A dimensional redefinition of the momentum components 
along the competing subspace has been performed (see the discussion following 
Eq.(1) in the last section). Since the resulting ``effective'' space dimension 
for the anisotropic cases is $d_{eff}= d -\frac{m}{2}$, where $d$ is the space 
dimension of the system, dimensional analysis yields
\begin{eqnarray}
\Gamma_{R (\tau)}^{(N)} (\rho_{\tau} k_{i (\tau)}, u_{\tau}, m_{\tau})&=&
\rho_{\tau}^{\tau(N + (d-\frac{m}{2}) - \frac{N(d-\frac{m}{2})}{2})}
\;\;\Gamma_{R (\tau)}^{(N)} (k_{i (\tau)}, u_{\tau}(\rho_{\tau}),
\frac{m_{\tau}}{\rho_{\tau}})\nonumber.
\end{eqnarray}

Now consider the asymptotic part of the corresponding vertex function 
satisfying the homogeneous CS equation. The solution is given by:
\begin{eqnarray}
\Gamma_{as\; R (\tau)}^{(N)} (k_{i (\tau)}, u_{\tau}, \frac{m_{\tau}}{\rho_{\tau}})&=&
exp[-\frac{N}{2} \int_{u_{\tau}}^{u_{\tau}(\rho_{\tau})} \gamma_{\phi (\tau)}(u'_{\tau}(\rho_{\tau}))
\frac{{d u'_{\tau}}}{\beta_{\tau}(u'_{\tau})}]\\
&&\Gamma_{as\; R (\tau)}^{(N)} (k_{i (\tau)}, u_{\tau}(\rho_{\tau}), 
m_{\tau})\nonumber, 
\end{eqnarray}
where 
\begin{equation}
\rho_{\tau}= \int_{u_\tau}^{u_{\tau}(\rho_{\tau})} \frac{du'_{\tau}}{\beta_{\tau}(u'_{\tau})} \; \; .
\end{equation}
From now on we shall drop the subscript for asymptotic in the vertex parts 
satisfying the homogeneous Callan-Symanzik equations. The values of the 
coupling constants $u_{\tau \infty}$ yielding the eigenvalue conditions 
$\beta_{\tau}(u_{\tau \infty})=0$ do exist, and precisely at this point 
the solutions to the CS equations under a scale in the external momenta 
obey the equations
\begin{eqnarray}
\Gamma_{R (\tau)}^{(N)} (\rho_{\tau} k_{i (\tau)}, u_{\tau \infty}, m_{\tau})&=&\rho_{\tau}^{\tau(N + (d-\frac{m}{2}) - \frac{N(d-\frac{m}{2})}{2})
-\frac{N \gamma_{\phi (\tau)}(u_{\tau \infty})}{2} } \\
&&\Gamma_{R (\tau)}^{(N)} (k_{i (\tau)}, u_{\tau \infty},m_{\tau})\nonumber .
\end{eqnarray}
The dimension of the field is defined by
\begin{equation}
\Gamma_{(\tau)}^{(N)}(\rho_{\tau}k_{i (\tau)}) = \rho_{\tau}^{\tau [(d-\frac{m}{2}) - N d_{\phi (\tau)}]}
\Gamma_{(\tau)}^{(N)}(k_{i}).
\end{equation}
This in turn implies that the definition of the anomalous dimension of the 
field through
$d_{\phi (\tau)}= \frac{d - \frac{m}{2}}{2} - 1 + \frac{\eta_{\tau}}{2 \tau}$ 
corresponds to $\eta_{\tau}=\gamma_{\phi (\tau)}(u_{\tau \infty})$. 

If we consider the vertex parts including composite operators, we can identify 
the anomalous dimension of the composite operator $\phi^{2}$ as follows. For 
$u_{\tau}=u_{\tau \infty}$ the aymptotic behavior of the vertex parts are 
given by $((N,L) \neq (0,2))$
\begin{eqnarray}
\Gamma_{R (\tau)}^{(N,L)} (\rho_{\tau} k_{i (\tau)},\rho_{\tau} p_{i (\tau)} , u_{\tau \infty}, m_{\tau})&=&\rho_{\tau}^{\tau(N + (d-\frac{m}{2}) - \frac{N(d-\frac{m}{2})}{2} -2L)-\frac{N\gamma_{\phi (\tau)}(u_{\tau \infty})}{2} + 
L\gamma_{\phi^{2} (\tau)}(u_{\tau \infty})} \\
&&\Gamma_{R (\tau)}^{(N)} (k_{i (\tau)}, p_{i (\tau)} , u_{\tau \infty},m_{\tau})\nonumber .
\end{eqnarray}
Writing the coefficient in the right hand side as 
$\rho_{\tau}^{\tau[(d-\frac{m}{2}) - N d_{\phi (\tau)}] + L d_{\phi^{2} (\tau)}}$, 
we conclude that $d_{\phi^{2} (\tau)} = -2 \tau + \gamma_{\phi^{2} (\tau)}(u_{\tau \infty})$. The 
correlation length exponents can now be obtained through the relation 
$\nu_{\tau}^{-1}= - d_{\phi^{2} (\tau)}= 2 \tau - \gamma_{\phi^{2} (\tau)}(u_{\tau \infty})$. We now have the resources to calculate these two exponents 
diagramatically.  

\subsection{Isotropic}

Let us briefly discuss the isotropic case $d=m$. Owing to the existence of 
only one type of scaling transformation, all we have to do is to concentrate 
on the competing sector ($\tau=2$) of the anisotropic treatment performed 
above. That means that we start with the mass scale $m_{3}$, dimensionless 
coupling constant $u_{3}$ and subscript 3 in all vertex functions in order to 
avoid confusion with the competing subspace of the anisotropic cases. There 
are 3 main differences. First, the critical dimension is 8. Therefore, the 
expansion parameter is $\epsilon_{L}= 8 - m$. Second, the effective space 
dimension is $\frac{m}{2}$. Finally, the beta function does not have the 
global factor of two as in the competition directions of the anisotropic case. 
Instead, in terms of dimensionless parameters one has 
$\beta_{3} = -  \epsilon_{L}(\frac{\partial ln u_{0 3}}{\partial u_{3}})^{-1}$. Following the same trend as before, we conclude that the anomalous dimension 
of the field in this case is 
$\eta_{L4}\equiv \eta_{3}= \gamma_{\phi (3)}(u_{3 \infty})$, where 
$u_{3 \infty}$ is the value of the coupling constant which yields 
$\beta_{3}(u_{3 \infty})=0$. In addition, the anomalous dimension of the 
composite operator is related to the correlation length exponent through the 
relation  
$\nu_{3}^{-1}= - d_{\phi^{2} (3)}= 4 - \gamma_{\phi^{2} (3)}(u_{3 \infty})$. Let us turn our attention to the calculation of these critical exponents by means of 
diagramatic expansions.

\section{ Anisotropic critical exponents}

We shall perform the computation of the critical exponents using the 
normalization conditions (previously introduced in Sec.II) for the anisotropic 
cases.

We start by using eqs.(2),(3) and (5) in order to determine the normalization 
functions $Z_{\phi (\tau)}, \bar{Z}_{\phi^{2} (\tau)}$ in powers of $u_{\tau}$ 
along with the Feynman integrals at the loop order required. We employ the 
orthogonal approximation in the calculation of Feynman integrals. 
The resulting expressions for the normalization functions 
can be written in terms of the four diagrams $I_{2}, I'_{3}, I_{4}, I'_{5}$ of 
one-, two- and three-loop order, respectively. They are presented in 
Appendix A. We recall that each loop integral produces a 
geometric angular factor $\frac{1}{4}S_m S_{d-m}
\Gamma(2-\frac{m}{4})\Gamma(\frac{m}{4})$, which shall be absorbed in a 
redefinition of the coupling constants. Performing those redefinitions, we 
obtain the renormalization functions in terms of the above mentioned diagrams, 
namely:
\begin{subequations}
\begin{eqnarray}
u_{0 \tau} =&& u_{\tau}[1 + \frac{(N+8)}{6} I_{2} u_{\tau}\nonumber \\ 
&& +(\frac{[(N+8)I_{2}]^{2}}{18} - (\frac{(N^{2}+6N+20)I_{2}^{2}}{36} + 
\frac{(5N+22)I_{4}}{9}) - \frac{(N+2)I'_{3}}{9})u_{\tau}^{2}],\\
Z_{\phi (\tau)} =&& 1 + \frac{(N+2)I'_{3}}{18} u_{\tau} 
+ \frac{(N+2)(N+8)(I_{2}I'_{3} - \frac{I'_{5}}{2})}{54}u_{\tau}^{2},\\
\bar{Z}_{\phi^{2} (\tau)} =&& 1 + \frac{(N+2)I_{2}}{6} u_{\tau}\nonumber \\
&& + [\frac{(N^{2}+7N+10)I_{2}}{18} - \frac{(N+2)}{6}(\frac{(N+2)I_{2}^{2}}{6} 
+ I_{4})]u_{\tau}^{2}.
\end{eqnarray}
\end{subequations}
As usual, the coefficients of the various terms in powers of $u_{\tau}$ have 
poles in $\epsilon_{L}$. These poles cancel in the calculation of the 
$\beta_{\tau}$ and the critical exponents. In fact using Eqs.(5), the 
$\beta_{\tau}$ and Wilson functions in either subspace are given by:
\begin{subequations}
\begin{eqnarray}
&& \beta_{\tau}  =  -\tau \epsilon_{L}u_{\tau}[1 - a_{1 \tau} u_{\tau}
+2(a_{1 \tau}^{2} -a_{2 \tau}) u_{\tau}^{2}],\\
&& \gamma_{\phi (\tau)} = -\tau \epsilon_{L}u_{\tau}[2b_{2 \tau} u_{\tau}
+ (3 b_{3 \tau}  - 2 b_{2 \tau} a_{1 \tau}) u_{\tau}^{2}],\\
&& \bar{\gamma}_{\phi^{2} (\tau)} = \tau \epsilon_{L}u_{\tau}[c_{1 \tau}
+ (2 c_{2 \tau}  - c_{1 \tau}^{2} - a_{1 \tau} c_{1 \tau})u_{\tau}    ].
\end{eqnarray}
\end{subequations}
Substitution of the explicit values of the coefficients given in (13) along 
with the results for the diagrams computed in Appendix A yields the following 
expression for $\beta_{\tau}$
\begin{eqnarray}
\beta_{\tau}=&& -\tau u_{\tau}[\epsilon_{L} 
- \frac{(N+8)}{6}(1+([i_{2}]_{m}-1))u_{\tau} \nonumber\\
&&  - \frac{(3N+14)}{12}u_{\tau}^{2}] + O(u_{\tau}^{4}).
\end{eqnarray}
From this expression, a zero of order $\epsilon_{L}$ of each 
$\beta_{\tau}$ characterizing the independent noncompeting ($\tau=1$) and 
competing ($\tau=2$) subspaces correspond to the same 
value of the coupling constant 
($u_{1 \infty}=u_{2 \infty}\equiv u_{\infty})$, i.e.,
\begin{equation}
u_{\infty}=\frac{6}{8 + N}\,\epsilon_L\Biggl\{1 + \epsilon_L
\,\Biggl[ - ([i_{2}]_m -1) + \frac{(9N + 42)}{(8 + N)^{2}}\Biggr]\Biggr\}\;\;.
\end{equation}   
Replacing this value back in the functions $\gamma_{\phi (\tau)}$ and 
$\bar{\gamma}_{\phi^{2} (\tau)}$ together with the coefficients given in 
(14) and in Appendix A, we find 
\begin{eqnarray}
&&\gamma_{\phi(\tau)}(u_{\infty})= \frac{\tau}{2} \epsilon_{L}^{2}\,\frac{N + 2}{(N+8)^2}
[1 + \epsilon_{L}(\frac{6(3N + 14)}{(N + 8)^{2}} - \frac{1}{4})] .
\end{eqnarray}
Notice that these expression are exactly the same of those for the 
critical exponents $\eta_{L2}\equiv \eta_{1}$ and $\eta_{L4}\equiv \eta_{2}$ 
obtained previously in \cite{Lei1} upon the identification 
$\gamma_{\phi(1)}(u_{\infty})=\eta_{L2}$ and 
$\gamma_{\phi(2)}(u_{\infty})=\eta_{L4}$.

The exponents $\nu_{\tau}$ can be found analogously from the expression of the 
anomalous dimension of the composite operator $\phi^{2}$. Recalling that 
$\gamma_{\phi^{2}(\tau)}= \bar{\gamma}_{\phi^{2}(\tau)}+ \gamma_{\phi (\tau)}$, it follows that 
$\nu_{\tau}^{-1}= - d_{\phi^{2} (\tau)}= 2 \tau 
- \gamma_{\phi^{2} (\tau)}(u_{\tau \infty}) 
- \gamma_{\phi(\tau)}(u_{\infty})$. Using the results in Appendix A once 
again, we obtain the following intermediate step:
\begin{equation}
\bar{\gamma}_{\phi^{2} (\tau)}(u_{\tau}) = \tau \frac{(N+2)}{6} u_{\tau}
[1 + \epsilon_{L}([i_{2}]_{m} - 1) - \frac{1}{2} u_{\tau}].
\end{equation}
Then, inserting the value $u_{\infty}$ into the last expression as well 
as $\gamma_{\phi(\tau)}(u_{\infty})$ from Eq.(18), we get 
\begin{eqnarray}
&& \nu_{\tau} =\frac{1}{2\tau} + \frac{(N + 2)}{4 \tau(N + 8)} \epsilon_{L}
+  \frac{1}{8 \tau}\frac{(N + 2)(N^{2} + 23N + 60)} {(N + 8)^3} \epsilon_{L}^{2}.
\end{eqnarray}
These exponents are in perfect agreement with those calculated 
previously using a massless theory \cite{Lei1} when we identify 
$\nu_{1}=\nu_{L2}$ and $\nu_{2} \equiv \nu_{L4}$, respectively. The other 
exponents can be obtained from these (computed diagrammatically) via scaling 
relations.

\section{Isotropic critical exponents}

We are going to use the main results derived in the previous section for the 
normalization functions in the presence of a single mass parameter and only 
one type of length/momentum scale corresponding to the isotropic critical 
behavior. In this case, each loop integral produces an angular factor of 
$S_{m}$, the area of an $m$-dimensional unit sphere, which shall be absorbed 
in a redefinition of the coupling constant. First, we describe the exponents 
using the orthogonal approximation for Feynman diagrams. Then, we shall 
present the results without the need for approximations in calculating the 
graphs, which is a simple feature of isotropic points.

\subsection{Exponents in the orthogonal approximation}  

The basic definitions of the dimensionless renormalized coupling constant 
and normalization functions can be retrieved from Eqs.(14) using the subscript 
$\tau=3$ in order not to confuse with the competing sector of the anisotropic 
cases. Moreover, the Feynman integrals appropriate for the isotropic case 
have to be used. The results of these diagrams using the orthogonal 
approximation are given in Appendix B, which are going to be used in this 
subsection. Furthermore, the functions obtained from the normalization 
constants and dimensionless coupling constant through Eqs.(6) in the isotropic 
cases now read
\begin{subequations}
\begin{eqnarray}
&& \beta_{3}  =  -\epsilon_{L}u_{3}[1 - a_{13} u_{3}
+2(a_{13}^{2} -a_{23}) u_{3}^{2}],\\
&& \gamma_{\phi (3)} = - \epsilon_{L}u_{3}[2b_{23} u_{3}
+ (3 b_{33}  - 2 b_{23} a_{13}) u_{3}^{2}],\\
&& \bar{\gamma}_{\phi^{2} (3)} = \epsilon_{L}u_{3}[c_{13}
+ (2 c_{23}  - c_{13}^{2} - a_{13} c_{13})u_{3}    ].
\end{eqnarray}
\end{subequations} 
Recall that in the above expressions $\epsilon_{L}=8-m$. Using the integrals 
presented in Appendix B, we find 
\begin{eqnarray}
\beta_{3}=&& - u_{3}[\epsilon_{L} 
- \frac{(N+8)}{6}(1 - \frac{1}{4}\epsilon_{L})u_{3} \nonumber\\
&&  + \frac{(3N+14)}{24}u_{3}^{2}] + O(u_{3}^{4}).
\end{eqnarray}
From the eigenvalue condition $\beta_{3}(u_{3 \infty})=0$ we obtain the 
following result for the zero of order $\epsilon_{L}$ of the dimensionless 
coupling constant
\begin{equation}
u_{3 \infty}=\frac{6}{8 + N}\,\epsilon_L\Biggl\{1 + \frac{\epsilon_L}{2}
\,\Biggl[ \frac{1}{2} + \frac{(9N + 42)}{(8 + N)^{2}}\Biggr]\Biggr\}\;\;.
\end{equation}
Using this value in the expression for $\gamma_{\phi (3)}$ in conjumination 
with Eqs.(14) and the results in  Appendix B, one can show that
\begin{eqnarray}
&&\gamma_{\phi(3)}(u_{3 \infty})= \epsilon_{L}^{2}\,\frac{N + 2}{4 (N+8)^2}
[1 + \epsilon_{L}(\frac{6(3N + 14)}{(N + 8)^{2}} - \frac{1}{8})] .
\end{eqnarray}
Once more, this is precisely the value of the exponent $\eta_{L4}$ previously 
obtained using the renormalization group equation in the massless 
theory \cite{Lei1}. Moreover, using again the results in Appendix B  
in the definition of $\bar{\gamma}_{\phi^{2} (3)}$, we obtain
\begin{equation}
\bar{\gamma}_{\phi^{2} (3)}(u_{3}) = \frac{(N+2)}{6} u_{3}
[1 -\frac{1}{4} \epsilon_{L} - \frac{1}{4} u_{3}].
\end{equation}
When Eq.(23) is replaced into this expression it yields
\begin{equation}
\bar{\gamma}_{\phi^{2} (3)}(u_{3 \infty}) = \frac{(N+2)}{(N+8)} \epsilon_{L}
[1 +\frac{(3N+9)}{(N+8)^{2}}\epsilon_{L} ],
\end{equation}
and recalling that 
$\nu_{3} \equiv \nu_{L4}=(4-\bar{\gamma}_{\phi^{2} (3)}(u_{3 \infty})-\gamma_{\phi(3)}(u_{3 \infty}))^{-1}$, we get to 
\begin{eqnarray}
&& \nu_{L4} =\frac{1}{4} + \frac{(N + 2)}{16(N + 8)} \epsilon_{L}
+  \frac{1}{256}\frac{(N + 2)(N^{2} + 23N + 60)} {(N + 8)^3} \epsilon_{L}^{2}.
\end{eqnarray}
The expression for $\bar{\gamma}_{\phi^{2} (3)}(u_{3 \infty})$ in Eq.(26) 
is the same as the one associated to a scalar theory in the massless limit, 
computed at the fixed point using the renormalization group equation. 
Besides, Eq.(27) corresponds to Eq.(208) for this exponent in the 
orthogonal approximation using the method of Ref.\cite{Lei1}. 

\subsection{Exponents in the exact computation}
The Feynman integrals of the isotropic case can be manipulated without 
referring to any approximation in the massive case, as described in 
Appendix C. Our goal here is to rederive the exponents using the machinery 
developed in the last subsection replacing, however, the expression for the 
diagrams by their exact counterparts presented in Appendix C.    

We start by writing down the $\beta_{3}$ function explicitly in terms of the 
diagrams calculated in Appendix C, which in that case reduces to  
\begin{eqnarray}
\beta_{3}=&& - u_{3}[\epsilon_{L} 
- \frac{(N+8)}{6}(1 - \frac{1}{4}\epsilon_{L})u_{3} \nonumber\\
&&  - \frac{(41N+202)}{1080}u_{3}^{2}] + O(u_{3}^{4}).
\end{eqnarray}
Therefore, this expression has a zero at the following value of the 
dimensionless coupling constant
\begin{equation}
u_{3 \infty}=\frac{6}{8 + N}\,\epsilon_L\Biggl\{1 + \frac{\epsilon_L}{2}
\,\Biggl[ \frac{1}{2} - \frac{(41N + 202)}{15(8 + N)^{2}}\Biggr]\Biggr\}\;\;.
\end{equation}
Following the same steps as before it is not difficult to show that this 
value of the coupling constant yields
\begin{eqnarray}
&&\gamma_{\phi(3)}(u_{3 \infty})= - \epsilon_{L}^{2}\,\frac{3(N + 2)}{20 (N+8)^2} + \frac{(N + 2)}{10(N + 8)^{2}}[\frac{(41N+202)}{10(N+8)^{2}} + \frac{23}{80}] .
\end{eqnarray}
This expression is equal to the value of the exponent $\eta_{L4}$ calculated 
exactly in conformity with that of $O(\epsilon_{L}^{2})$ firsty found by 
Hornreich, Luban and Shtrikman \cite{Ho-Lu-Sh} and recently confirmed and 
extended to $O(\epsilon_{L}^{3})$ in Ref.\cite{Lei2} for the massless case, 
as expected.

Finally, let us determine $\nu_{L4}$. First, we have to make use of the 
results in Appendix C to write down the following expression for arbitrary 
dimensionless coupling constant
\begin{equation}
\bar{\gamma}_{\phi^{2} (3)}(u_{3}) = \frac{(N+2)}{6} u_{3}
[1 -\frac{1}{4} \epsilon_{L} + \frac{1}{12} u_{3}].
\end{equation}
Second, we replace the especial value of the coupling constant 
$u_{3 \infty}$ in this equation in order to show that
\begin{equation}
\bar{\gamma}_{\phi^{2} (3)}(u_{3 \infty}) = \frac{(N+2)}{(N+8)} \epsilon_{L}
[1 - \frac{(13N+41)}{15(N+8)^{2}}\epsilon_{L} ],
\end{equation}
which allows us to figure out the value of $\nu_{L4}$ from 
$\gamma_{\phi(3)}(u_{3 \infty})$ and 
$\bar{\gamma}_{\phi^{2} (3)}(u_{3 \infty})$, namely
\begin{eqnarray}
&& \nu_{L4} =\frac{1}{4} + \frac{(N + 2)}{16(N + 8)} \epsilon_{L}
+  \frac{(N + 2)(15N^{2} + 89N + 4)} {960(N + 8)^3} \epsilon_{L}^{2}.
\end{eqnarray}
This is precisely the exponent $\nu_{L4}$ in the exact calculation 
recently found at $O(\epsilon_{L}^{2}$ \cite{Lei2}. As before, the 
other exponents can be discovered through the use of scaling relations.

\section{Discussion of the results and conclusions}

It is interesting to find two different methods producing the same values of 
the critical exponents calculated diagrammatically. Those 
previously computed using a massless scalar field theory subtracted at 
nonzero external momenta using the renormalization group equations 
in Refs.\cite{Lei1} are rederived herein using massive theories (with 
two or one independent mass scales depending whether we treat the anisotropic 
or isotropic cases) in the Callan-Symanzik method. 

Most of the remarks done originally by Brezin, Le Guillou and Zinn-Justin in 
Ref.\cite{BLZ2} for conventional critical phenomena are still valid here, for 
isotropic as well as anisotropic cases. First, the eigenvalue condition 
$\beta_{\tau}(u_{\tau \infty})=0$ corresponds to a 
scale invariant theory, as shown here, but it is not attractive: the theory 
is only scale invariant precisely at this point, since the derivative
$\beta'_{\tau}(u_{\tau \infty})$ calculated at this point is positive.
Second, our calculations are valid in the region of momenta much larger 
than the mass in any particular subspace. The difference dwells in the 
anisotropic cases: two inequivalent subspaces require two independent mass 
scales along with two coupling constants which consistently have the same 
eigenvalue leading to scale invariance separately in each subspace. 
Universality is recovered when the renormalized dimensionless coupling 
constants are fixed at the solution of the eigenvalue $u_{\tau \infty}$. 

One step further within the formalism introduced in the present work is to 
extend it in order to treat arbitrary competing systems, when arbitrary types 
of competing directions characterized by higher order derivative terms of 
arbitrary powers are involved\cite{Lei2}, generalizing the 
present concept. The Feynman integrals get much more difficult to solve, 
especially when one tries to perform the exact calculation for the isotropic 
cases. That is the reason we have calculated here the integrals using the 
orthogonal approximation and the exact computation for those integrals. In 
spite of yielding different values for the critical exponents, the difference 
among them in each calculation is negligible as already pointed out in 
\cite{Lei2}. We shall leave the subject of generic competing systems in the 
Callan-Symanzik method for future work \cite{Lei3}.  

The anisotropic behavior with two independent mass scales and the analysis 
developed herein might be useful in addressing further anisotropic problems 
in quantum field theory, possibly with broken Lorentz/Poincar\'e invariance.
First, it is useful to consider the higher order terms in momentum space 
appearing isotropically in the ultraviolet regime. This possibility arises, 
for instance, in Lorentz-violating dispersion relations for cosmic ray 
processes due to the nontrivial character of the short-distance structure of 
spacetime \cite{Amelino}. This effect can be implemented in the 
propagation of massive particles as well, even anisotropically. One 
mathematical possibility to describe such phenomena is that the metric tensor 
depends upon the momentum variables. This has been done recently to 
investigate the simplest example in anisotropic Planck-scale modified 
dispersion relations for a relativistic particle of mass $m$ propagating in 
2 dimensions taking into account phenomenological quantum gravitational 
effects (i.e, the modified dispersion relations) in the framework of Finsler 
geometry. The coefficient of the cubic term in the momentum $p_{1}$ in the 
mass-shell condition is directly related to the Planck mass 
$M$ \cite{Girelli}. In principle, these modified dispersion 
relations can be postulated for massive fields as well. Then the procedure 
should be the opposite to that described in the present paper to obtain the 
Lagrangian: keep $\delta_{0} \neq 0$, restore the 
dimensionful character of the higher order derivative terms and introduce 
the equations of motion with any power of momenta (mass-shell condition) 
isotropically or anisotropically, as a constraint using an auxiliary field 
(Lagrange multiplier). The problem is that the equation of motion obtained 
by varying the Lagrange multiplier is not easy to solve (even in the particle 
case) and the renormalization group structure in this framework is not 
obvious. 

Another perspective is to follow the trend proposed here of 
maintaining $\delta_{0}= 0$ which allows the decoupling of the inequivalent 
subspaces and makes it easier the ultraviolet renormalization of quantum 
fields. Siegel's recipe to introduce masses by dimensional reduction in 
Poincar\'e invariant field theories \cite{Siegel} would be modified in the 
present case with at least two inequivalent subspaces. This would require two 
extra dimensions to characterize completely the two mass scales involved in 
the problem, one for each momentum subspace. This is so because one can 
mathematically formulate a gradient operator for ``Lifshitz spaces'' which 
can be defined as $\partial_{i} \equiv (\partial_{1},...,\partial_{(d-m)},
\partial_{new (d-m+1)} \equiv  \partial^{2}_{old (d-m+1)},..., 
\partial_{new (d)} \equiv \partial^{2}_{old (d)})$ through the dimensional 
redefinition proposed in \cite{Lei1}, with the same Euclidean metric, keeping 
in mind that the mass $m_{1}$ characterizes the noncompeting (quadratic) 
subspace, while $m_{2}$ is the typical scale of the competing (quartic) 
subspace, with the latter having half of the canonical dimension of the 
former. This new operator would reflect directly the ``folding'' of the 
dimensions along the competing subspace as a dynamical effect provoked by 
the combination of attractive and repulsive components of the background on 
which quantum fields propagate. This new flat space limit does not correspond 
to Minkowski space. Although this is trivial for scalar fields, it is 
definitely not the case for defining spinors and vector fields in these 
Lifshitz spaces with anisotropic higher derivative terms in the kinetic part.

In conclusion, the new method proposed in the present work to study $m$-axial 
Lifshitz points for the massive scalar field theory renormalized at zero 
external momenta opens the possibility to study field theories with 
inequivalent subspaces characterized by two (or more) mass scales. 
Universality is confirmed, since the critical exponents calculated in the 
massive theory using the Callan-Symanzik formalism agree with those obtained 
in the massless scalar field theory renormalized at nonvanishing external 
momenta.  

\section{Acknowledgments}

We would like to acknowledge CNPq for financial support (PRSC) and 
partial financial support (MML). 

\appendix 
\section{Feynman graphs for anisotropic behaviors}

We compute the appropriate diagrams corresponding to one-, two- and three-loop 
integrals using dimensional regularization and Feynman parameters to fold 
several denominators together. Only the one-loop integral for the four-point 
function can be solved exactly for anisotropic criticalities. The higher loop 
diagrams can be calculated, however, using the orthogonal approximation, which 
is the most general approximation consistent with the physical property of 
homogeneity. The basic statement of this approximation is that the loop 
momenta in a given subdiagram is orthogonal to all loop momenta appearing in 
other subdiagrams. Due to our normalization conditions defined in the text, 
we shall work with zero external momenta. Explicitly, the integrals we have 
to resolve are given by:

\begin{equation}
I_2 =  \int \frac{d^{d-m}q d^{m}k}{[\bigl((k)^{2}\bigr)^2 +
(q)^{2} + 1]^{2}}\;\;\;,
\end{equation}
$ I_{3}'= \frac{\partial I_{3}(P,K)}{\partial P^{2}}|_{P=K=0}(= 
\frac{\partial I_{3}(P,K)}{\partial K^{4}}|_{P=K=0})$ where $I_{3}(P,K)$ 
is the integral
\begin{equation}
I_{3} = \int \frac{d^{d-m}{q_{1}}d^{d-m}q_{2}d^{m}k_{1}d^{m}k_{2}}
{\left( q_{1}^{2} + (k_{1}^{2})^2 + 1 \right)
\left( q_{2}^{2} + (k_{2}^{2})^2 + 1 \right)
[(q_{1} + q_{2} + P)^{2} + \bigl((k_{1} + k_{2}+ K)^{2}\bigr)^2 + 1]}\;\;,
\end{equation}

\begin{eqnarray}
I_{4}\;\; =&& \int \frac{d^{d-m}{q_{1}}d^{d-m}q_{2}d^{m}k_{1}d^{m}k_{2}}
{\left( q_{1}^{2} + (k_{1}^{2})^2 + 1 \right)^{2}
\left( q_{2}^{2} + (k_{2}^{2})^2 + 1  \right)}\nonumber\\
&&\qquad\qquad\qquad \times \frac{1}
{[(q_{1} + q_{2})^{2} + \bigl((k_{1} + k_{2})^{2}\bigl)^2 +1]}\;\;,
\end{eqnarray}
and $ I_{5}'= \frac{\partial I_{5}(P,K)}{\partial P^{2}}|_{P=K=0}(= 
\frac{\partial I_{5}(P,K)}{\partial K^{4}}|_{P=K=0})$ where $I_{5}(P,K)$ 
is the three-loop diagram 

\begin{eqnarray}
I_{5}\;\; =&&
\int \frac{d^{d-m}{q_{1}}d^{d-m}q_{2}d^{d-m}q_{3}d^{m}k_{1}d^{m}k_{2}
d^{m}k_{3}}
{\left( (q_{1}+P)^{2} + ((k_{1}+K)^{2})^2 + 1 \right)
\left( q_{2}^{2} + (k_{2}^{2})^2 + 1 \right)
\left( q_{3}^{2} + (k_{3}^{2})^2 + 1 \right)}\nonumber\\
&&\times\frac{1}{[ (q_{1} + q_{2})^{2} 
+ \bigl((k_{1} + k_{2})^{2}\bigr)^2 + 1 ]
[(q_{1} + q_{3})^{2} + \bigl((k_{1} + k_{3})^{2}\bigr)^2 + 1]}.
\end{eqnarray}
Proceeding as indicated above, their $\epsilon_{L}$-expansion have the 
following expressions 
\begin{equation}
I_{2} = \frac{1}{\epsilon_{L}} [ 1 + ([i_{2}]_{m}-1)\epsilon_{L}] + O(\epsilon_{L}),
\end{equation}
\begin{equation}
I_{3}^{'} = \frac{-1}{8 \epsilon_{L}}
[1+(2[i_{2}]_{m} - \frac{5}{4}) \epsilon_{L}] - \frac{1}{8}I 
+ O(\epsilon_{L}), 
\end{equation}
\begin{equation}
I_{4} = \frac{1}{2 \epsilon_{L}^{2}} \Bigl(1 +
(2\;[i_{2}]_m  -\frac{3}{2}) \epsilon_{L} + O(\epsilon_{L}^{2})\Bigr),
\end{equation}
\begin{equation}
I_{5}^{'} =
\frac{-1}{6 \epsilon_{L}^{2}}
[1+(3[i_{2}]_{m} -\frac{7}{4})\epsilon_{L}+ O(\epsilon_{L}^{2} ] - \frac{1}{4\epsilon_{L}} I,
\end{equation}
where $[i_{2}]_{m}= 1 + \frac{1}{2}(\psi(1)-\psi(2-\frac{m}{4}))$ and 
\begin{equation}
I= \int_{0}^{1}dx (\frac{1}{1-x(1-x)} + \frac{ln[x(1-x)]}{[1-x(1-x)]^{2}}) ,
\end{equation}
is the same integral appearing in the original work by BLZ. This integral 
disappears in the calculation of the critical exponents. Its appearance here 
is a general feature of the orthogonal approximation, since it will be present 
in the final result of the calculation of the diagrams in the isotropic cases 
as well. It does not appear in the masless theory and is only an artifact of 
our choice of the normalization point for the external momenta.

\section{Isotropic diagrams in the generalized orthogonal approximation}
We shall now turn our attention to the isotropic integrals in an expansion in 
$\epsilon_{L}=8-d$. The simplifying feature of this case is the existence of 
only one type of correlation length, which implies only one mass as usual in 
field theory calculations. The issue here is a quartic power in the propagator. The expressions for these graphs are:
\begin{equation}
I_2 =  \int \frac{d^{m}k}{[\bigl((k)^{2}\bigr)^2 + 1]^{2}}\;\;\;,
\end{equation}
$ I_{3}'= \frac{\partial I_{3}(P,K)}{\partial K^{4}}|_{K=0}$ where 
$I_{3}(K)$ 
is the integral
\begin{equation}
I_{3} = \int \frac{d^{m}k_{1}d^{m}k_{2}}
{\left((k_{1}^{2})^2 + 1 \right)
\left((k_{2}^{2})^2 + 1 \right)
[\bigl((k_{1} + k_{2}+ K)^{2}\bigr)^2 + 1]}\;\;,
\end{equation}

\begin{eqnarray}
I_{4}\;\; =&& \int \frac{d^{m}k_{1}d^{m}k_{2}}
{\left( (k_{1}^{2})^2 + 1 \right)^{2}
\left( (k_{2}^{2})^2 + 1  \right)}\nonumber\\
&&\qquad\qquad\qquad \times \frac{1}
{[\bigl((k_{1} + k_{2})^{2}\bigl)^2 +1]}\;\;,
\end{eqnarray}
and $ I_{5}'= \frac{\partial I_{5}(K)}{\partial K^{4}}|_{K=0}$ where 
$I_{5}(K)$ 
is the three-loop diagram 

\begin{eqnarray}
I_{5}\;\; =&&
\int \frac{d^{m}k_{1}d^{m}k_{2}
d^{m}k_{3}}
{\left( ((k_{1}+K)^{2})^2 + 1 \right)
\left( (k_{2}^{2})^2 + 1 \right)
\left( (k_{3}^{2})^2 + 1 \right)}\nonumber\\
&&\times\frac{1}{[\bigl((k_{1} + k_{2})^{2}\bigr)^2 + 1 ]
[\bigl((k_{1} + k_{3})^{2}\bigr)^2 + 1]}.
\end{eqnarray}
Using the same steps as indicated in the orthogonal approximation for the 
anisotropic behaviors we find for their $\epsilon_{L}$-expansion the 
following results   
\begin{equation}
I_{2} = \frac{1}{\epsilon_{L}} [ 1 - \frac{1}{4} \epsilon_{L}] + O(\epsilon_{L}),
\end{equation}
\begin{equation}
I_{3}^{'} = \frac{-1}{16 \epsilon_{L}}
[1 - \frac{1}{8} \epsilon_{L}] - \frac{1}{32}I 
+ O(\epsilon_{L}), 
\end{equation}
\begin{equation}
I_{4} = \frac{1}{2 \epsilon_{L}^{2}} \Bigl(1 - \frac{1}{4}\epsilon_{L} + O(\epsilon_{L}^{2})\Bigr),
\end{equation}
\begin{equation}
I_{5}^{'} =
\frac{-1}{12 \epsilon_{L}^{2}}
[1 - \frac{1}{8}\epsilon_{L}+ O(\epsilon_{L}^{2}) ] - \frac{1}{16\epsilon_{L}} I,
\end{equation}
where $I$ is the same integral which appears in Appendix A. As before this 
integral do not contribute for the critical exponents.   

\section{Isotropic integrals in the exact calculation}
We shall consider now the integrals without performing approximations. The 
expressions are the same of those given in the previous appendix. Before 
quoting the results some commentaries are in order. First, the simplicity of 
the isotropic cases allow us to perform the integrals exactly as advertised. 
A useful trick is to write the propagators in the form 
$\frac{1}{[(k^{2})^{2} + 1]}= \frac{1}{(k^{2} + i)(k^{2}-i)}$ and then use a 
Feynman parameter. One can proceed as before since now the propagators are 
quadratic in the loop momenta, but the resulting parametric integrals turns 
out to be more complicated than those in the orthogonal approximation. 
Moreover, some care must be exercised in order to manipulate the remaining 
factors of $i$. Setting $\epsilon_{L}=0$ in the powers of $i$ when performing 
the parametric integrals effectively suppress it. We can then figure out the 
final expressions for the exact calculation, namely
\begin{equation}
I_{2} = \frac{1}{\epsilon_{L}} [ 1 - \frac{1}{4} \epsilon_{L}] + O(\epsilon_{L}),
\end{equation}
\begin{equation}
I_{3}^{'} = \frac{3}{80 \epsilon_{L}}
[1 - \frac{77}{120} \epsilon_{L}] - \frac{9}{2}H 
+ O(\epsilon_{L}), 
\end{equation}
\begin{equation}
I_{4} = \frac{1}{2 \epsilon_{L}^{2}} \Bigl(1 - \frac{7}{12}\epsilon_{L} + O(\epsilon_{L}^{2})\Bigr),
\end{equation}
\begin{equation}
I_{5}^{'} =
\frac{1}{20 \epsilon_{L}^{2}}
[1 - \frac{13}{15} \epsilon_{L}+ O(\epsilon_{L}^{2}) ] 
- \frac{9}{\epsilon_{L}} H,
\end{equation} 
and now the multiparametric integral is specific to $m$-axial isotropic 
Lifshitz points and turns out to be 
\begin{eqnarray}
H=&& \int_{0}^{1}dz z(1-z)\int_{0}^{1}\int_{0}^{1}\int_{0}^{1}dxdydt\int_{0}^{1}du u(1-u)^{3} \nonumber\\
&& \times\;\; ln \Bigl(u[\frac{2x(1-z)+2yz-1}{z(1-z)} -2t+1] +2t-1\Bigr).
\end{eqnarray}
The integral H (as well as the integral I appearing in the orthogonal 
approximation) does not contribute to the critical exponents since it 
disappears in the renormalization algorithm.

\end{document}